# Probing a Self-Developed Aesthetics Measurement Application (SDA) in Measuring Aesthetics of Mandarin Learning Web Page Interfaces

*Jasni Mohamad Zain[†], Mengkar Tey[††], and Yingsoon Goh[†††]*

[†] and [††]*Faculty of Computer Systems and Software Engineering, University Pahang Malaysia, Karung Berkunci 12, 25000 Kuantan, Pahang, Malaysia.*
[†††]*Academy of Language Studies, UiTM Terengganu, 23000 Dungun, Terengganu, Malaysia.*

**Summary**

This article describes the accurateness of our application namely Self-Developed Aesthetics Measurement Application (SDA) in measuring the aesthetics aspect by comparing the results of our application and users' perceptions in measuring the aesthetics of the web page interfaces. For this research, the positions of objects, images element and texts element are defined as objects in a web page interface. Mandarin learning web pages are used in this research. These learning web pages comprised of main pages, learning pages and exercise pages, on the first author's E-portfolio web site. The objects of the web pages were manipulated in order to produce the desired aesthetic values. The six aesthetics related elements used are balance, equilibrium, symmetry, sequence, rhythm, as well as order and complexity. Results from the research showed that the ranking of the aesthetics values of the web page interfaces measured of the users were congruent with the expected perceptions of our designed Mandarin learning web page interfaces (reported also in [18]). Thus, it implies that the subjectivity of aesthetics can be measured in an objective manner. For further understanding on our effort in doing aesthetics measurement, we discussed in details the object modelling used as well as the process in developing this application. We also explained the steps on how to run our application. Additionally, the strength and the weakness of our SDA application shared in this paper suggest that there is still room for the improvement for aesthetics measurement.

## 1. Introduction

This paper reviews the current state visual aesthetics measurement for the Mandarin learning web page interfaces particularly and for web page interfaces in general. Specific focuses are given on the discussion on the importance of visual aesthetics measurement and the ways to develop an aesthetics-measuring application to measure the aesthetics of the Mandarin web page interfaces. We thus suggest an appropriate approach on how to measure the aesthetics of web page interfaces in [17]. We have developed and used a simple application to measure the aesthetics of the Mandarin learning web page interfaces.

Aesthetics of web pages refers to how attractive a web page is in which it catches the attention of the user to read the information. Visual appearance is important in getting attentions of the users to browse through the entire web pages. Parizotto-Ribeiro *et al.* [15] stated that those screens perceived as aesthetically pleasing were having a better usability. In addition, aesthetics measuring methods of web page interface are important as this may help in gaining users' attention and in erecting their interest in using the interface. However, aesthetics of the web page interfaces can be very subjective, because different people might have their own dissimilar views.

Therefore, it is normally very difficult to judge whether an interface is exquisite or not. Consequently, the major concern of this research is to provide an objective tool for unbiased aesthetics measurement. Some research showed that an important aspect of screen design is aesthetics evaluation of screen layouts. A very essential component of GUI design entails the actual layout of elements on the screen. Web page interface aesthetics measuring methods are important to web page developers as this may help then in assuring the gaining of users' attention as well as in attracting their interest in assessing the web pages.

In order to achieve the above-mentioned purpose, Mandarin learning web pages used for this research comprised of main pages, learning pages and exercise pages that were manipulated according to the desired aesthetic values. We incorporated balance, equilibrium, symmetry, sequence, rhythm, as well as order and complexity our calculation based on Ngo *et al.*'s model [13].

The result of the findings showed that the ranking from the users' perceptions were remained the same as the measurements using the Self-Developed Aesthetics Measurement Application (SDA) in [17]. It was found that the users' perceptions were congruent with the aesthetics values gathered by using our Self-Developed Aesthetics Measurement Application (SDA). Hence, our SDA perhaps can be introduced as an effortless tool for web page aesthetics measurement.

For this paper, we concentrate the discussion on the strength of this SDA, weakness of the SDA, and some problems faced in the process of developing this SDA. In addition, we share our approach on how we developed this application and how the aesthetic values of the Mandarin learning web page interfaces were calculated.

## 2. Literature Reviews

Aesthetics of the web page interface grasps researchers' attention in the process of webpage interface design. Garrett [3] said that the visual design of the web pages referred to the balance, emotional appeal, or aesthetic of the web pages. Efforts have been carried out for aesthetics measurements. Approaches and ways of measuring aesthetics were conducted and the calculations and technique of aesthetics measurement for web page interface were also derived from the past research.

Ngo *et al.* [13] did a fundamental research focused on measuring aesthetic value for graphic screens. They discovered fourteen aesthetic measurements related elements, which were balance, equilibrium, symmetry, sequence, cohesion, unity, proportion, simplicity, density, regularity, economy, homogeneity, rhythm, as well as order and complexity. Their empirical study has suggested that these elements are very important in gaining users' attention to use the created computer systems. Ngo and Law [14] described their design principles and grouped them into five categories, which were, spatial relations of objects, directions of objects, proximity of objects, dimensions of objects, and locations of objects. While Kim *et al.* [9] along with Park *et al.* [16] identified design factors related to web-page objects, backgrounds as well as the relationships between the objects and the backgrounds.

Prior studies on the aesthetic aspects of interfaces done by Kim and Moon [8], focused on visual design elements as the objects. The visual design elements of a graphical user interface (GUI) regarded as objects would have effect on the aesthetic quality of a GUI. Faria and Oliveira [2] added on some of the aesthetics measures elements, which included general score, colours, page cover, text font family and font size, horizontal and vertical distortion, horizontal position, horizontal position, etc. By giving examples of aesthetic measures applied, they proposed their evaluative differences of the aesthetic values between the document template and its instances. Helen [6] presented her exact and accurate mathematical proof of symmetry measures where symmetry was proven as one of the most fundamental principles in design. This is because symmetrical page gives a feeling of permanence and stability.

Several design factors have also been associated with overall assessment of web-page attractiveness as mentioned by Lindgaard *et al.* [11]. A practical measure of document layout aesthetics included alignment, regularity, separation, balance, white-space fraction, white-space free flow, proportion, uniformity, and page security, have been described by Harrington *et al.* [5]. Their approach was to collect and combine heuristic rules for factors that can harm the aesthetics. Besides, the aesthetics measure of balance was well understood by both experts and non-experts. Its preference in visual displays was well documented by Wilson and Chatterjee [20]. Lavie and Tractinsky [10] developed specific questionnaires to evaluate aesthetics with components of classical aesthetics and expressive aesthetics. The classical aesthetics refereed to clean and symmetrical design and the expressive aesthetics refereed to creative and fascinating design, as well as usability.

Parizotto-Ribeiro *et al.* [15] found that there was a positive correlation between the aesthetics aspects (unity, proportion, homogeneity, balance, and rhythm) and perceived usability of VLEsU interfaces. In the other word, those screens perceived as aesthetically pleasing, were having a better usability. In addition, Toh [19] supported that the good aesthetic layouts definitely affected a student's motivation to learn, as related to ARCS model proposed by Keller and Suzuki [7]. This ARCS model referred to Attention, Relevance, Confidence, and Satisfaction, as shown in Table 1. In relation to this, Tey *et al.* [17] discovered that aesthetically pleasing layouts of web page would motivate students in Mandarin learning. Besides, Goh *et al.* [4] asserted that the attractiveness of an e-portfolio would be a pulling factor that drawn students in the use of e-portfolio.

All these latest literature reviews showed that aesthetics is an important aspect of interface design. In order to attract users' use of the interfaces as well as to increase the usability of any interfaces, aesthetics should not be neglected. Thus, the measurement of aesthetics is a crucial task that needs to be worked out.

Table 1: ARCS Model

| A | Attention | Good layouts will attract the attention of the student. |
|---|---|---|
| R | Relevance | Good layouts will be relevant to the student. |
| C | Confidence | Good layouts will boost the student's confidence. |
| S | Satisfaction | The student will feel satisfied if the design is good and appealing. |

## 3. Approach and Methodology

In order to measure aesthetics values, we use our Self-Developed Aesthetics Measurement Application (SDA). Objects used for our aesthetics measuring purpose referred to the positions of objects, images element and texts element in several Mandarin learning web page interfaces. These Mandarin learning web pages were uploaded to the first author's E-portfolio. These Mandarin Learning web pages were developed, designed, and manipulated according to the desired aesthetics values. A brief description of our SDA was discussed below.

### 3.1 Self-Developed Aesthetics Measurement Application (SDA)

This SDA was developed by using Matlab software based on six elements derived from the model of Ngo *et al.* [13]. It was developed by using image-processing method, MATLAB Graphical User Interface development environment (GUIDE) and other necessary functions equipped in Matlab software.

Figure 1 showed the main interface of our SDA. There were several buttons placed at the right of this GUI Aesthetics Measurer. To start measuring GUI's aesthetics, the users may click on the first button: Choose interface. Then, the user of this application may click the button of Count Aesthetics Value. The application would automatically do all required calculations by running the formulae set which we discussed in detail in the next sub-session below.

Further information would be provided by clicking on the "information" button. By clicking on the button of "close application", the user would then exit from our application. If the user needed to know more detail, they may click the button of "Show Detail". The details included number of objects that were dragged, width, height, area, difference between objects and frames, centre point of objects, etc.

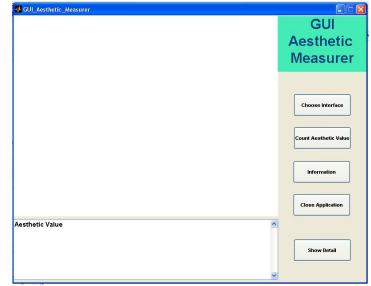

Figure 1: Main Interface of Our SDA

### 3.2 Formulae Involved in the SDA

Table 2 showed the explanations of terms used and the formulae used for aesthetics measurements. There were altogether six major components used for aesthetics measurement. The six elements involved were balance, equilibrium, symmetry, sequence, rhythm, as well as order and complexity. The definitions of terms as well as the mathematical formulae were included.

Table 2: Six Major Components Used for Aesthetics Measurement
Source: Ngo, Teo, and Byrne (2003)

| 1. Balance |
|---|
| $Balance = 1 - \frac{|BalanceVer| + |BalanceHor|}{2} \in [0,1]$ |
| Balance was computed as the difference between total weighting of components on each side of the horizontal and vertical axis. |
| 2. Equilibrium |
| $Equilibrium = 1 - \frac{|Equilibrium_X| + |Equilibrium_Y|}{2} \in [0,1]$ |
| Equilibrium was computed as the difference between the center of mass of the displayed elements and the physical center of the screen. |
| 3. Symmetry |
| $Symmetry = 1 - \frac{|SymmetryVer| + |SymmetryHor| + |SymmetryRad|}{3} \in [0,1]$ |
| Symmetry was the extent to which the screen is symmetrical in three directions: vertical, horizontal, and diagonal. |
| 4. Sequence |
| $Sequence = 1 - \frac{\sum_{j=UL,UR,LL,LR} |q_j - v_j|}{8} \in [0,1]$ |
| Sequence was a measure of how information in a display was ordered in relation to a reading pattern that was common. |
| 5. Rhythm |
| $Rhythm = 1 - \frac{|Rhythm_X| + |Rhythm_Y| + |Rhythm_{Area}|}{3} \in [0,1]$ |
| Rhythm was the extent to which the objects are systematically ordered. |
| 6. Order and Complexity |
| $Order\_Complexity = \frac{\sum_{i}^{5} M_i}{5} \in [0,1]$ |
| The measure of order was written as the sum of the above measures for a layout. The opposite pole on the continuum was complexity. The scale created might also be considered as a scale of complexity, with extreme complexity at one end and minimal complexity (order) at the other. |

3.3 Mandarin Learning Web Page Interfaces

Table 3 showed the twelve Mandarin learning web page interfaces. These web pages were developed and designed according to the required aesthetic values. They were altogether four main pages, four learning pages and four exercise pages. They were arranged into four groups, which were labeled as Main Page (Group 1), Main Page (Group 2), Main Page (Group 3), and Main Page (Group 4), Learning Page (Group 1), Learning Page (Group 2), and subsequently. Group 1 was Mandarin web page interfaces with highest aesthetics values, while Group 4 was Mandarin web page interfaces with the lowest aesthetics values. Table 4 showed the object models of the twelve Mandarin learning web pages. These object models were manipulated by using image-processing method. Photoshop program was used for this purpose. They were used as object models of the web page interfaces. The purpose of this object modeling was to show the objects on each Mandarin learning web page interface clearly.

Table 3: Twelve Mandarin Learning Web Page Interfaces

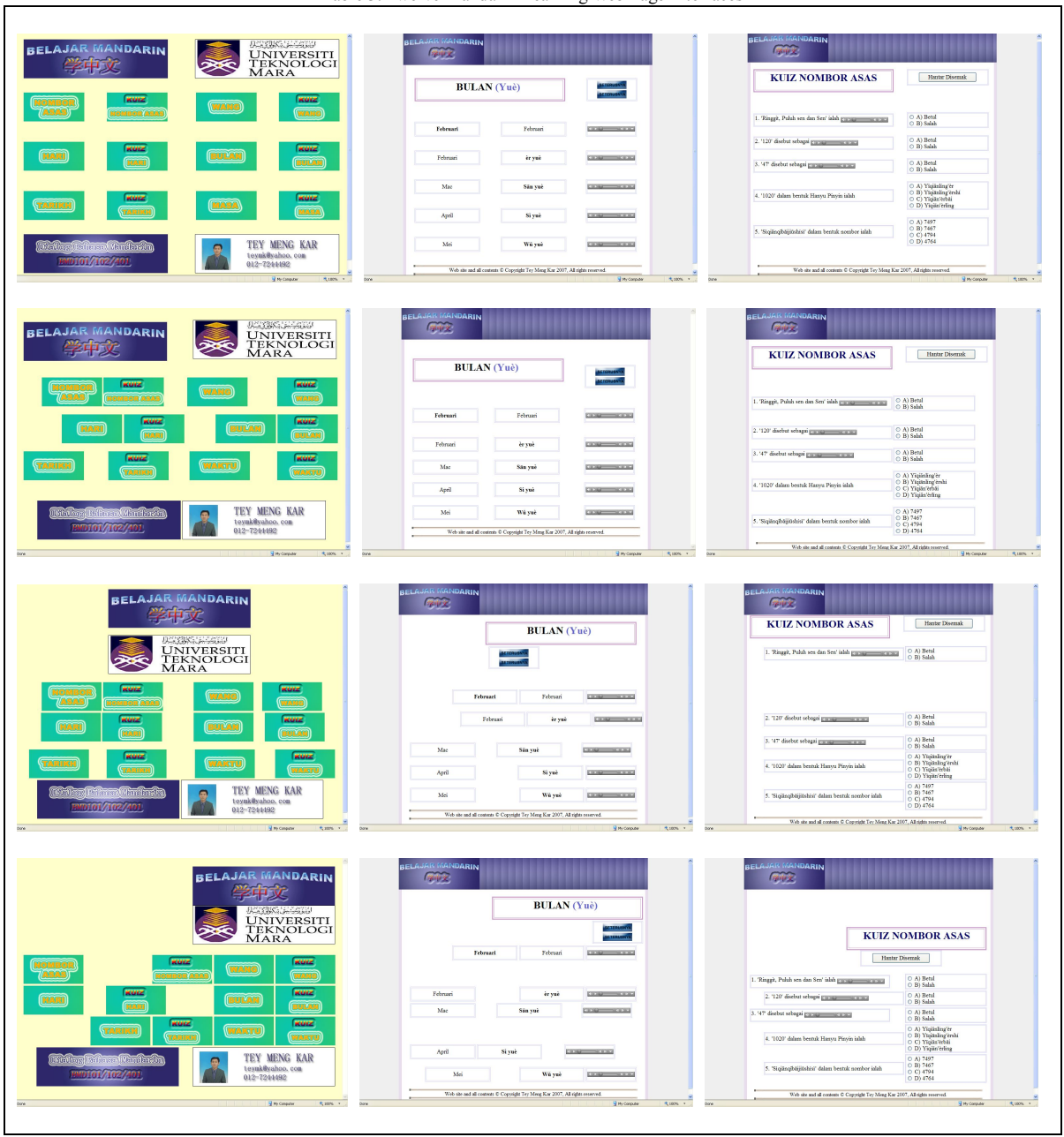

Table 4: Object Model of Twelve Mandarin Learning Web Page Interfaces

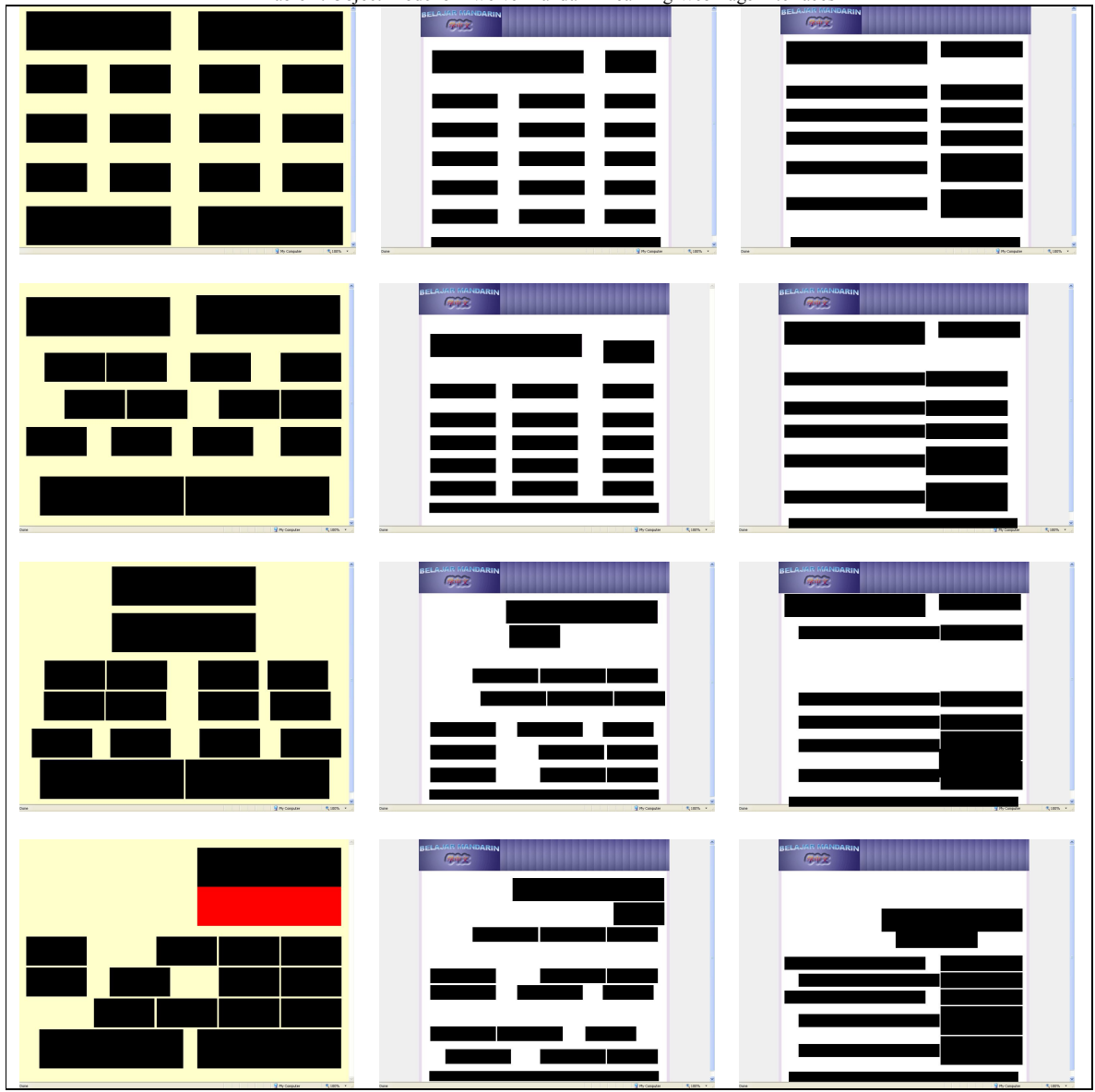

## 3.3 How to Use Our Application

Below were the steps in using our application for aesthetics measurements. In order to give a clearer picture, we used Main Page (Group 1) for discussion purpose:
 i. Opened the SDA application;
 ii. Clicked 'Choose Interface" button to select a web page interface that to be measured;
 iii. Dragged all objects on the interface as shown in Figure 4;
 iv. Clicked "Count Aesthetic Value" button to get the aesthetic value of interface after all objects were dragged;
 v. The aesthetics value of the interface of a particular web page would be shown at the mini command window. The values showed included Balance, Equilibrium, Symmetry, Sequence, Rhythm, Order and Complexity as well as the overall Aesthetics Value (OM). Figure 5 showed the aesthetics values counted for this Main page interface;

vi. All the information of objects dragged and aesthetics values were saved in a directory of which the SDA application execute file was located. This information could be seen by clicking the button "Show Detail". Figure 6 showed the information of objects dragged and all the aesthetics values. This information included number of objects, width, height, area, difference between objects and frames, centre point of objects while the aesthetics values showed were Balance, Equilibrium, Symmetry, Sequence, Rhythm, Order and Complexity as well as the overall Aesthetics Value (OM).

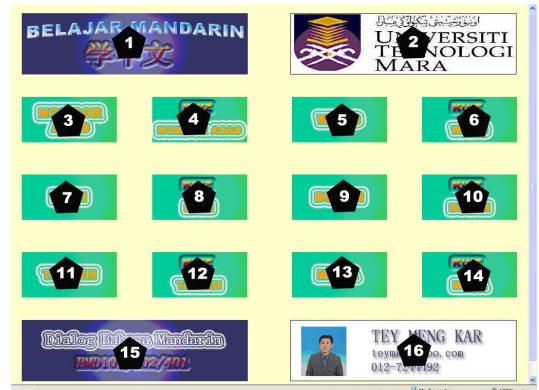

Figure 4: Object Model of Main Page (Group 1) with Objects Numbered

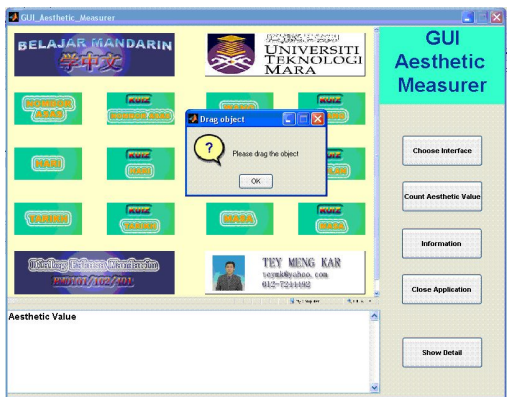

Figure 2: Execution Dialogue: Drag Objects on Interface

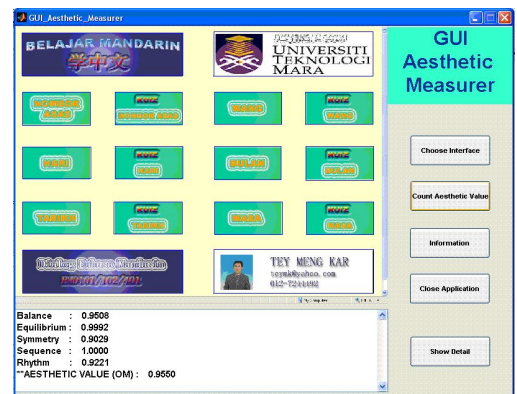

Figure 5: Aesthetic Value of Main Page Interface

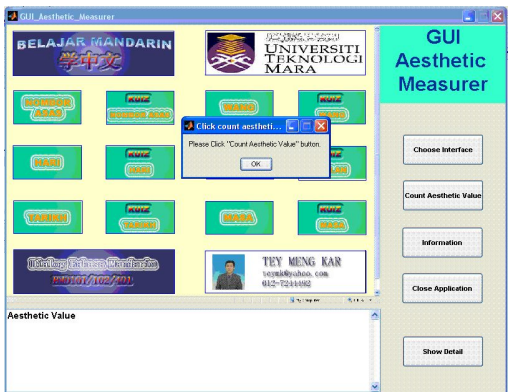

Figure 3: Execution Dialogue: Click "Count Aesthetic Value" Button

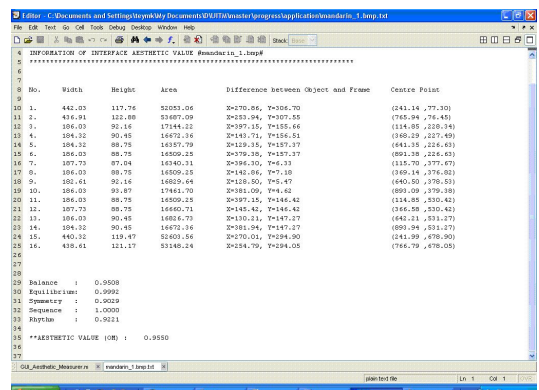

Figure 6: Information of Objects Dragged and Aesthetic Values of Main Page (Group 1)

As a concrete example of how we used this application for our research, Table 5 showed the results of all of the aesthetics values of twelve web pages used in our research. These values were between 0 (the worst) and 1 (the best).

Table 5: Results of Aesthetic Values (avs) of Mandarin Learning Web Pages by Using SDA

| Main Page (Group 1) | 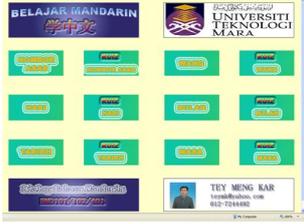 | Learning Page (Group 1) | 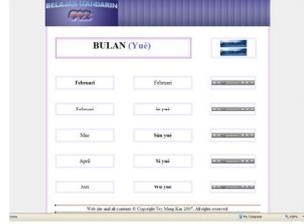 | Exercise Page (Group 1) | 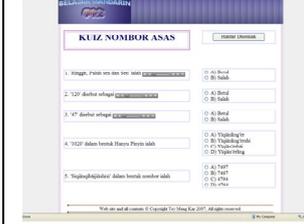 |
|---|---|---|---|---|---|
| Balance | 0.9445 | Balance | 0.6558 | Balance | 0.8054 |
| Equilibrium | 0.9991 | Equilibrium | 0.9954 | Equilibrium | 0.9965 |
| Symmetry | 0.9013 | Symmetry | 0.6062 | Symmetry | 0.4402 |
| Sequence | 1.0000 | Sequence | 0.7500 | Sequence | 0.7500 |
| Rhythm | 0.9085 | Rhythm | 0.6663 | Rhythm | 0.5592 |
| **Aesthetic value (av)** | **0.9507** | **Aesthetic value (av)** | **0.7347** | **Aesthetic value (av)** | **0.7103** |
| Main Page (Group 2) | 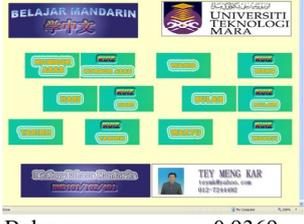 | Learning Page (Group 2) | 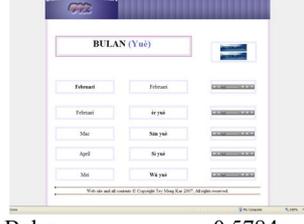 | Exercise Page (Group 2) | 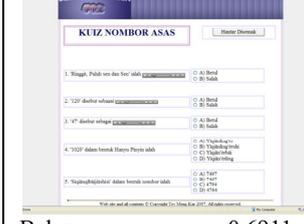 |
| Balance | 0.9369 | Balance | 0.5784 | Balance | 0.6911 |
| Equilibrium | 0.9990 | Equilibrium | 0.9945 | Equilibrium | 0.9932 |
| Symmetry | 0.8234 | Symmetry | 0.4161 | Symmetry | 0.3796 |
| Sequence | 1.0000 | Sequence | 0.7500 | Sequence | 0.7500 |
| Rhythm | 0.8700 | Rhythm | 0.4917 | Rhythm | 0.4331 |
| **Aesthetic value (av)** | **0.9259** | **Aesthetic value (av)** | **0.6461** | **Aesthetic value (av)** | **0.6494** |
| Main Page (Group 3) | 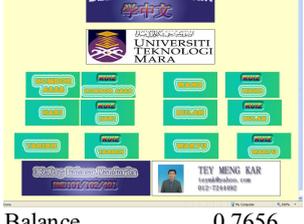 | Learning Page (Group 3) | 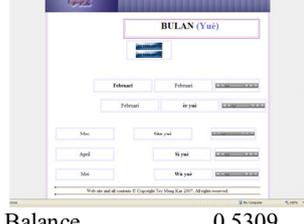 | Exercise Page (Group 3) | 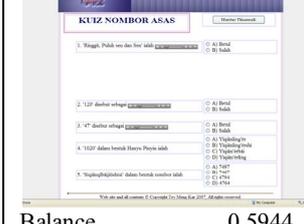 |
| Balance | 0.7656 | Balance | 0.5309 | Balance | 0.5944 |
| Equilibrium | 0.9960 | Equilibrium | 0.9935 | Equilibrium | 0.9913 |
| Symmetry | 0.4958 | Symmetry | 0.4555 | Symmetry | 0.4515 |
| Sequence | 0.6250 | Sequence | 0.5000 | Sequence | 0.5000 |
| Rhythm | 0.5324 | Rhythm | 0.4870 | Rhythm | 0.3459 |
| **Aesthetic value (av)** | **0.6830** | **Aesthetic value (av)** | **0.5934** | **Aesthetic value (av)** | **0.5766** |
| Main Page (Group 4) | 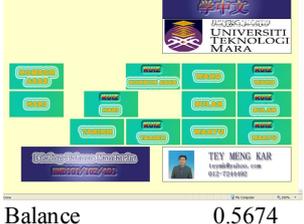 | Learning Page (Group 4) | 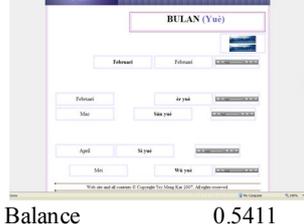 | Exercise Page (Group 4) | 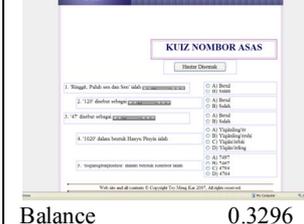 |
| Balance | 0.5674 | Balance | 0.5411 | Balance | 0.3296 |
| Equilibrium | 0.9918 | Equilibrium | 0.9934 | Equilibrium | 0.9859 |
| Symmetry | 0.2689 | Symmetry | 0.3399 | Symmetry | 0.3421 |
| Sequence | 0.3750 | Sequence | 0.3750 | Sequence | 0.5000 |
| Rhythm | 0.2258 | Rhythm | 0.3511 | Rhythm | 0.3134 |
| **Aesthetic value (av)** | **0.4858** | **Aesthetic value (av)** | **0.5201** | **Aesthetic value (av)** | **0.4942** |

## 4. Findings and Discussions

Table 6 showed the comparisons between the rankings of results using Self-Developed Aesthetics Measurement Application (SDA). The results of the application showed that web pages of group 1 were the most aesthetic and followed by group 2, group 3, and the group 4 was the most unaesthetic ones (reported also in [18]).

Table 6: Comparison between Ranking of the Results of Application

| GROUP | Mandarin Learning Web page Interface | | | | | |
|---|---|---|---|---|---|---|
| | Main Page | | Learning Page | | Exercise Page | |
| | Value | Rank | Value | Rank | Value | Rank |
| 1 | 0.9507 | 1st | 0.7347 | 1st | 0.7103 | 1st |
| 2 | 0.9259 | 2nd | 0.6461 | 2nd | 0.6494 | 2nd |
| 3 | 0.6830 | 3rd | 0.5934 | 3rd | 0.5766 | 3rd |
| 4 | 0.4858 | 4th | 0.5201 | 4th | 0.4942 | 4th |

The ranking of all values in Group 1 stayed at the first position for all aspects; therefore, our application reflected users' views on the aesthetics aspect of the web pages. In unison, the ranking of all the values in Group 4 remained the last position in all aspects, thus, it also supported that our application did in accordance to users' perceptions on the aesthetics aspect of the web pages. However, the aesthetics values of the Main pages in Group 1 and Group 2 were not very much distinctly different in terms of their aesthetics values measured. It might be due to the shortcomings in the designs of Main pages in Group 1 and Group 2.

As a whole, the results were congruent with the initial designed aesthetics values of Mandarin learning web page interfaces. Therefore, our SDA perhaps could be introduced as an effortless tool for web page aesthetics measurement. There is some strength found in our SDA. Our SDA application can be used to calculate aesthetics values easily. The users may just need to start the application and run the process of dragging objects. All these can be done by only having several clicks and they will be able to get the needed aesthetics values.

Furthermore, this application is also very simple to use. There are not any complicated steps that need to be followed. The users also do not need to have any pre-requisite knowledge or expertise to make the application run for them. The main user interface is very much user friendly.

However, there are some weaknesses of our SDA. One of the weaknesses was regarding the ways our application recognizes the objects of the interface. The first time user might not know what the "objects" of their interface are. This would then cause the inaccuracy of dragging the appropriate selected object on their web page interface. This would thus trigger inaccurate aesthetics values. Likewise, the users need to drag objects from their interface. To inexperience users, dragging objects from interface might not be easy though the objects allowed in this application limited to regular shapes only, such as squares, circles, etc. This limitation is also an obvious weakness of this application. Consequently, this application might be applicable to the other irregular shapes, which are much in prevalent use in the interfaces nowadays.

There are some problems faced during the development process of SDA. One of the essential problems was in the process of choosing appropriate elements for aesthetics measurement. There are abundant of elements found. Thus, it was not possible to include all those mathematical formulae into the Matlab software in order to produce a powerful application. We have to choose the six very crucial ones. By abandoning some other applicable elements, off course, this will sure cut down the correctness of our application.

On the other hand, we also faced some problems in designing the suitable Mandarin learning web page interfaces for our experimental studies purpose. We need to produce web pages with significant differences in aesthetic values. These web pages need to be developed, designed, and manipulated according to the desired aesthetic measurements. It is quite challenging to design web pages of different aesthetics values but with similar contents.

## 5. Implications, Suggestions, and Conclusion

The findings of the results showed that the calculations and technique of aesthetics measurement in this research were as expected. Hence, our SDA perhaps can be introduced as an effortless tool for web page aesthetics measurement.

Taking as a whole, our Self-Developed Aesthetics Measurement Application (SDA) can still be considered as a viable tool in evaluating the aesthetics of a web page interface. What can be improved is that, there were only six aesthetic elements, which consisted of balance, equilibrium, symmetry, sequence, rhythm, as well as order and complexity used to develop this application. Consequently, other influential factors might not have been taken into account. Thus, more aesthetics measurers should be entailed in developing a more powerful web page interface aesthetics-measuring tool.

On the other hand, more emphasizes should be practiced to improve the weakness of the SDA discussed above. Automated calculation function is suggested where by the users do not need to drag of any objects on a particular interface or an object-recognizable tool should be devised to ease the users. By using more powerful image processing and recognizing method, and with more works and research done on this are, it is hoped that a simple and useful tool can be produced for evaluating the aesthetics aspects of the web page interface.

Furthermore, shapes of objects, colours of objects, texts, frames, and background of interface are also very vital and should not be ignored. This research can also be extended to the measuring of aesthetics of multi windows, multi document, and multi pane interfaces.

Aesthetic responses are multi-dimensional and not just limited to the single dimension of beauty. It could include qualities such as 'adorable,' 'cool,' or 'strong,' as well, as mentioned by Liu [12]. Besides, several web pages should be analyzed to determine the common aesthetic design factors and the corresponding emotional responses of users, as mentioned by Kim *et al.* [9]. Therefore, this research can also be extended by focusing on aesthetics impressions and corresponding emotional responses of users.

In conclusion, measurement of aesthetics can be viewed as an important task that should not be disregarded. This is because aesthetics of interface would influence usability, acceptability, learn-ability, comprehensibility, and productivity. Thus, the subjectivity of aesthetics should be measured in an objective manner to produce desirable outcomes of the use of any interfaces. Therefore, there is still room for the improvement for aesthetics measurement.

# Reference

[1] Agrawal, R., Imielinski, T. and Swami, A. 1993. Mining association rules between sets of items in large databases. Proc. ACM SIGMOD International Conference on Management of Data, Washington DC, USA, **22**:207-216, ACM Press.

[2] Faria, A. C. and Oliveira, J. B. S. de. 2006. Measuring Aesthetic Distance between Document Templates and Instances. Proceedings of the 2006 ACM symposium on Document engineering. 13-21.

[3] Garrett, J. J. 2003. *The Elements of User Experience: User-centered Design for the Web*. New Riders Publications, Indiana.

[4] Goh, Y. S., Tey, M. K., and, Jasni, M. Z. 2007. E-Portfolio to Supplement the Teaching of Mandarin as a Foreign Language. Proceedings of the 5th International Conference on Information Technology in Asia (CITA 2007). Sarawak, Malaysia.

[5] Harrington, S. J., Naveda, J. F., and Jones, R. P. 2004. Aesthetic Measures for Automated Document Layout. Proceedings of the 2004 ACM symposium on Document engineering. 109-111.

[6] Helen, B. 2006. Evaluating interface aesthetics: A measure of symmetry. Proceedings of the SPIE, **6076**: 52-63.

[7] Keller, J. M. and Suzuki, K. 1988. Use of the ARCS motivation model in courseware design. *Instructional Designs for Microcomputer Courseware*, Lawrence Erlbaum Associates, Hillsdale NJ. 401-434.

[8] Kim, J. and Moon, J. Y. 1998. Designing towards aesthetic usability in customer interface. *Interacting with Computers*, **10 (1)**: 1-29.

[9] Kim, J., Lee, J. and Choi, D. 2003. Designing emotionally evocative homepages: an empirical study of the quantitative relations between design factors and emotional dimensions. *International Journal of Human–Computer Studies*. **59 (6)**: 899-940.

[10] Lavie, T., Tractinsky, N. 2004. Assessing dimensions of perceived visual aesthetics of web sites. *International Journal of Human–Computer Studies*. **60 (3)**: 269-298.

[11] Lindgaard, G., Fernandes, G. J., Dudek, C., Brownet, J. 2006. Attention web designers: you have 50 ms to make a good first impression!. *Behaviour and Information Technology*. **25 (2)**: 115-126.

[12] Liu, Y. 2001. Engineering Aesthetics and Aesthetic Ergonomics: A Dual-process Methodology and its Applications. Proceedings of the International Conference on Affective Human Factors Design. *Asean Academic Press*, London. 248-255.

[13] Ngo, D. C. L., Teo, L. S., and Byrne, J. G. 2003. Modeling interface aesthetics. *Journal of Information Sciences*. **152 (1)**: 25-46.

[14] Ngo, D. C. L. and Law, B. L. 2003. An expert screen design and evaluation assistant that uses knowledge-based backtracking. *Information and Software Technology*. **45**: 293-304.

[15] Parizotto-Ribeiro, R., Hammond, N., Mansano, J., and Cziulik, C. 2004. Aesthetics and perceived usability of VLEs: Preliminary results. Proceedings of Human Computer Interaction. 217-221.

[16] Park, S., Choi, D. and Kim, J. 2004. Critical factors for the aesthetic fidelity of web pages: empirical studies with professional web designers and users. *Interacting with Computers*. **16**: 351-376.

[17] Tey, M. K., Goh, Y. S. and, Jasni, M. Z. 2007. Aesthetics of Multi Screen Interface and Its Relevance with Mandarin Learning. Proceedings of the 3rd International Conference on Open and Online Learning (ICOOL 2007). Penang, Malaysia.

[18] Tey, M. K., Jasni, M. Z. and Goh, Y. S. 2007. Measuring Aesthetics of Mandarin Learning Web Pages: Are Users' Perceptions Congruent with the Measured Values of Aesthetics-Measurement Application (AMA)? Proceedings of the National Conference on Software Engineering and Computer Systems (NaCSES 2007). Pahang, Malaysia.

[19] Toh, S. C. 1998. Cognitive and motivational effects of two multimedia simulation presentation modes on science learning. Ph.D. Dissertation. University of Science Malaysia, Malaysia.

[20] Wilson, A., Chatterjee, A. 2005. The assessment of preference for balance: introducing a new test. *Empirical Studies of the Arts*. **23 (2)**: 165-180.

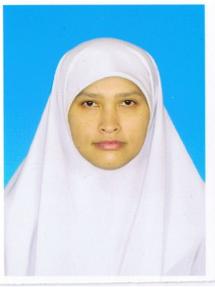

**Jasni Mohamad Zain** received her Bachelor degree in Computer Science from University of Liverpool, England, UK in 1989; PGCE Mathematics from Sheffield Hallam University, England, UK in 1994; M.E. degree from Hull University, England, UK in 1998 and PhD from Brunel University, West London, UK in 2005. She currently holds the post as the Director of the Center of Information Technology and Communication, University Malaysia Pahang. She is currently a lecturer in Faculty of Computer Science and Software Engineering, University Malaysia Pahang. She has been actively presenting papers in national and international conferences. Her research interests include Image Processing as well as Data and Network Security.

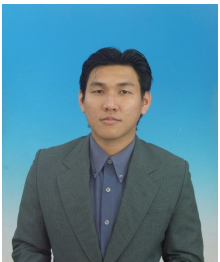

**Mengkar Tey** currently is pursuing his master degree in Software Engineering in University Malaysia Pahang under supervision Dr. Jasni Mohamad Zain and Mr. Yingsoon Goh. He has experiences in teaching Mandarin as the third language to non-native learners in MARA University of Technology, Malaysia for almost 2 years. His research interests are on the web page development, image processing, as well as Chinese character handwriting recognition.

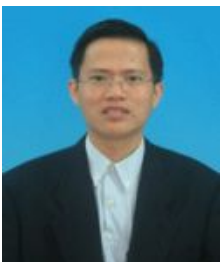

**Yingsoon Goh** currently teaches Mandarin as the third language to non-native learners in MARA University of Technology, Malaysia. He has experiences in teaching Mandarin at primary, secondary, and tertiary level for almost 17 years. He has been actively presenting papers in national and international conferences. His research interests are on the use of educational technology in Mandarin teaching and learning, as well as web-based instruction